\documentclass[useAMS,usenatbib]{mn2e}
\usepackage{graphicx}
\usepackage{color}

\title[Arp147]
  {SWIFT Observations of the Arp 147 Ring galaxy system.}
\author[L. Fogarty et al.]
  {L.~Fogarty,$^1$\thanks{lmrf@astro.ox.ac.uk}
  N.~Thatte,$^1$ M.~Tecza,$^1$ F.~Clarke,$^1$ T.~Goodsall,$^{1,2}$  R.~Houghton,$^1$ 
  \newauthor
   G.~Salter,$^1$ R.L.~Davies,$^1$ S. A.~Kassin$^1$
\\
$^1$Sub-Department of Astrophysics, University of Oxford, Denys Wilkinson Bldg., Keble Road, Oxford OX1 3RH U.K.
\\        
$^2$Jet Propulsion Laboratory, California Institute of Technology, 4800 Oak Grove Drive, MS 169-327, Pasadena, CA 91109, USA}

\date{Released 2002 Xxxxx XX}

\pagerange{\pageref{firstpage}--\pageref{lastpage}} \pubyear{2010}

\begin{document}

\label{firstpage}

\maketitle

\begin{abstract}

We present observations of Arp 147, a galaxy system comprising a collisionally-created ring galaxy and an early-type galaxy, using the Oxford SWIFT integral field spectrograph (IFS) at the 200-inch Hale telescope. We derive spatially resolved kinematics from the IFS data and use these to study the interaction between the two galaxies. We find the edge-to-edge expansion velocity of the ring is $225\pm8\rm{\,km\,s^{-1}}$, implying an upper limit on the timescale for the collision of 50\,Myrs. We also calculate that the angle of impact for the collision is between $33^{\circ}-\,54^{\circ}$, where 0$^{\circ}$ would imply a perpendicular collision. The ring galaxy is strongly star-forming with the star formation likely to have been triggered by the collision between the two galaxies.

We also measure some key physical parameters in an integrated and spatially resolved manner for the ring galaxy. Using the observed B$-$I colours and the H$\alpha$ equivalent widths, we conclude that two stellar components (a young and an old population) are required everywhere in the ring to simultaneously match both observed quantities. We are able to constrain the age range, light and mass fractions of the young star formation activity in the ring, finding a modest age range, a light fraction of less than a third, and a negligible ($<$1\%) mass fraction. We postulate that the redder colours observed in the south-east corner of the ring galaxy could correspond to the nuclear bulge of the original disk galaxy from which the ring was created, consistent with the stellar mass in the south-east quadrant being 30--50\% of the total. The ring appears to have been a typical disk galaxy prior to the encounter.

The ring shows electron densities consistent with typical values for star-forming HII regions. The eastern half of the ring exhibits a metallicity a factor of $\sim2$ higher than the western half. The ionisation parameter, measured across the ring, roughly follows the previously observed trend with metallicity.

\end{abstract}


\section{Introduction}
\subsection{Ring Galaxies and Arp 147}

Ring galaxies are a rare type of galaxy commonly observed with a nearby companion. Examples of such objects include the Cartwheel galaxy \citep{Fosbury77, Toomre1978, DaviesMorton82, MAH92}, II Hz 4 \citep{LyndsToomre76, TheysSpiegel76, AppletonMarston97, Romano08}, VII Zw 466 \citep{TheysSpiegel76, AppletonMarston97, Romano08} and AM 064-741 \citep{FewMadoreArp82}, to name a few that have been studied extensively.

Arp 147 is a ring galaxy with an early-type companion from the Arp catalogue of peculiar galaxies \citep{Arp66}. Taking  $H_{0}=73\pm3\rm{\,km\,s^{-1}Mpc}^{-1}$ \citep{Spergel07} and given the measured systemic velocity of $9656\pm18\rm{\,km\,s^{-1}}$ \citep{deVaucouleurs91} the system is at a distance of $132\,{\rm Mpc}$ implying a scale of $0.64\,{\rm kpc}$ per arcsecond. Some basic information for Arp 147 is shown in Table \ref{tab:basic}.

Most ring galaxies are thought to be produced by a collision between an intruder galaxy and a normal disk galaxy, as first put forward by \cite{LyndsToomre76} and \cite{TheysSpiegel77}. In this scenario the intruder galaxy collides with a normal disk galaxy causing a gravitational disturbance in the disk. As the intruder passes through the disk material is pulled towards the centre of the disk (i.e. the position of the intruder). The material then rebounds outwards once the intruder has passed through the disk, causing an over density of material in the form of an outwards propagating ring.

The basic parameters governing the geometry of the collision are the impact parameter and the angle of impact. The impact parameter describes how close the centres of the colliding galaxies pass each other, measured as the percentage of the disk radius between the point of impact and the centre of the disk. The angle of impact measures how far from perpendicular the collision is. Therefore in a purely symmetric collision, where the galaxies meet each other head on and perpendicularly, the impact parameter is 0 and the impact angle is 0$^{\circ}$

The symmetric scenario was modelled very successfully by \cite{Mihos94} who examined an interaction between a disk of stars and gas in a dark matter halo and an intruding gasless system. Examining a head-on, perpendicular collision between these two objects they were able to recreate a system very much like the Cartwheel galaxy, with the nucleus of the original disk remaining in place. These authors also modelled the star formation triggered in the ring finding that the maximum star formation enhancement occurs late in the interaction, some time after the companion has passed though the ring. 

It was noted, however, by \citet{LyndsToomre76} and \citet{Toomre1978} that this highly symmetric collision can seem contrived. In fact these authors suggest that varying the angle of impact by up to 30$^{\circ}$ to normal can still yield ring-like shapes, also suggesting a modest impact parameter (up to 15\% of the disk radius) will likewise still produce ring-like structures. The variation of these parameters was studied in more detail by \citet{HuangStewart88} who modelled collisions with a combination of various impact angles and impact parameters. They found that variations of the impact angle of up to 30$^{\circ}$ show little deviation from a perpendicular impact with the nucleus of the disk not significantly dislodged. Larger impact angles and the introduction of a non-zero impact parameter, on the other hand, can cause the nucleus to become merged with or even offset from the resulting ring. These authors also conclude that an impact angle of 45$^{\circ}$ is most disruptive to the disk and that arc-like structures can still be formed for impact parameters up to about 50\% of the disk length. 

In the context of Arp 147 \citet{Gerber92} investigated an off-centre, perpendicular collision between systems similar to those used by \cite{Mihos94}. They recreated a system like that of Arp 147, identifying the redder region in the south-east of the ring as the nucleus  and predicting it to be displaced relative to the plane of the ring. The IFS observations presented in this paper allow us to study the kinematics of the system in a spatially resolved manner and to examine the parameters of the collision that created the ring. 

It was predicted by \citet{TheysSpiegel76} that the ring is a transient phenomenon on a time-scale of $\sim10^{8}\rm{\,yrs}$. Since star formation is expected to be triggered by the expanding ring, there should be observational evidence of enhanced star formation in the ring and the star forming areas should have similar colours and ages. This makes ring galaxies very attractive to study since their star formation history should demonstrate a single spatially contiguous starburst around the ring.

\cite{AppletonMarston97} found that ring galaxies are generally blue and show a radial colour gradient in the sense that the colours become bluer in the outer part of the ring. This reinforces the idea that the star formation occurs in the ring and as the ring propagates outwards the stars age in its wake. \cite{Bransford98} also showed that star-forming knots seen in most ring galaxies tend to have very similar colours, further suggesting that they have similar ages and were formed in the ring at the same time. \cite{MarstonAppleton95} found that ring galaxies have similar H$\alpha$ luminosities to starburst galaxies, confirming the results of \cite{AppletonSM87_1} who found that ring galaxies have far infrared luminosities brighter than normal galaxies and comparable starburst galaxies. These authors also found that the star formation is almost exclusively found in the ring itself and not inside it.

Some kinematic studies of ring galaxies have also been carried out. The pioneering study by \cite{TheysSpiegel76} presented kinematics for two galaxies, VII Zw 466 and Arp 147. For the latter system they were able to derive only tentative velocities which none the less were shown to be suggestive of expansion across the ring. For VII Zw 466 the authors were able to derive expansion and rotation velocities for the ring by using data from many slit spectra taken at different positions across the ring. \cite{Fosbury77} carried out similar work for the Cartwheel galaxy, deriving rotation and expansion velocities for the ring. Other studies include \cite{FewMadoreArp82} who looked at slit spectra of AM 064-741 and \cite{TaylorAtherton84} who studied the Vela ring galaxy using a Fabry-Perot interferometer. \cite{Appleton96} presented some sample velocity maps from a ring galaxy simulation, which could be compared to observed two-dimensional kinematic maps. 

It is very clear that three-dimensional IFS data sets yielding two-dimensional kinematic maps of ring galaxies would be a rich addition to the field. The work presented in this paper provides an example of this.

Some recent studies in various wavebands have focussed on Arp 147. \cite{Rappaport10} looked at muliti-wavelength images of the system, including infra-red and X-ray data. They found a star formation rate in the ring consistent with $7\rm{\,M_{\odot}yr^{-1}}$, and derived an interaction timescale of $\sim40\rm{\,Myrs}$. They also located several ultra-luminous X-ray sources associated with the star-forming ring. 

\cite{Romano08} also looked at Arp 147, deriving a star-formation rate of $8.6\rm{\,M_{\odot}yr^{-1}}$ from their measured far infra-red flux. From their measurements of the combined H$\alpha$ and [NII] flux they derive a star formation rate and equivalent width slightly lower than ours. \cite{AppletonSM87_1} derived a far infrared luminosity for Arp 147, corresponding to a SFR of  $3.7\rm{\,M_{\odot}yr^{-1}}$ according to the relation of \cite{Kennicutt98}. The star formation rates derived in these studies are broadly consistent with ours, given the scatter in the various relations used to derive the rates.

The Arp 147 system, as well as being a fascinating object in its own right, is also an excellent example of an unfilled ring with a displaced nucleus. Several systems like this are known and an insight into their nature may result in a better understanding of the formation of these intriguing systems.

\subsection{The SWIFT Integral Field Spectrograph (IFS)}
\label{sec:SW}

The SWIFT Integral Field Spectrograph (IFS) is an instrument built by a dedicated team at the University of Oxford \citep{Thatte06}. SWIFT is an AO-assisted, slicer-based IFS operating in the I \& z bands. The instrument is mounted at the Cassegrain focus of the 200-inch Hale telescope at Palomar Observatory in California, where it was installed and commissioned in October 2008. SWIFT operates with the Palomar Adaptive Optics system (PALAO) in three modes - using the laser guide star (LGS-AO), using a natural guide star (NGS-AO), or in static correction mode. The latter puts a fixed pattern on the deformable mirror to remove static aberrations in the telescope and AO system and provides no further correction. SWIFT has an on-the-fly selectable spatial pixel (spaxel) scale consisting of coarse, intermediate and fine spaxels (0\farcs{235}, 0\farcs{160}, 0\farcs{080} respectively). The instrument field-of-view is 44$\times$89 spaxels in size yielding an on-sky field-of-view of 10\farcs{3}$\times$20\farcs{9}, 7\farcs{0}$\times$14\farcs{2}, 3\farcs{5}$\times$7\farcs{1} in the three spaxel scales. The instrument records $\sim$\,4000 spectra in each exposure. There are two selectable wavelength ranges of 650nm--1050nm or 750nm--1050nm which are selected by changing a dichroic element in the PALAO system. The instrument has a twin spectrograph design such that the field of view is split into two pseudo-slits at the image slicer. Each pseudo-slit then passes through a separate but identical spectrograph and is imaged onto one of two CCD detectors. SWIFT uses two 4k$\times$2k, 250$\mu$m thick, fully depleted, CCD arrays developed by LBNL \citep{LBNL}. These devices have very high quantum efficiency up to $\sim$\,1$\mu$m, even when compared to deep depletion devices \citep{Burke04}. 

\section{The Data}
\subsection{SWIFT Observations}
Arp 147 was observed with the SWIFT IFS on 12th January 2009. The instrument was operated in coarse spaxel mode, with 0\farcs{235} spaxels,  giving a field of view of 10\farcs{3}$\times$20\farcs{9}. The wavelength range was 650nm--1050nm covering the region of the H$\alpha$; [NII] $\lambda\lambda$ 6548, 6583~\AA; [SII] $\lambda\lambda$ 6716, 6731~\AA; and [SIII] $\lambda\lambda$ 9069, 9532~\AA~emission lines at the redshift of this object.

\begin{table}
\begin{tabular}{c|c}
\hline
\multicolumn{2}{c}{\bf{Arp 147}} \\
\hline
RA (h:m:s J2000) & 03:11:18.9 \\
Dec ($^\circ$:':" J2000) & 01:18:53 \\
Redshift & 0.032 \\
Systemic Velocity ($\rm{km\,s^{-1}}$) & $9656\pm18$ \\
\hline
\end{tabular}
\caption{Arp 147 Basic Information}
\label{tab:basic}
\end{table}

\begin{figure}
\includegraphics[width=84mm]{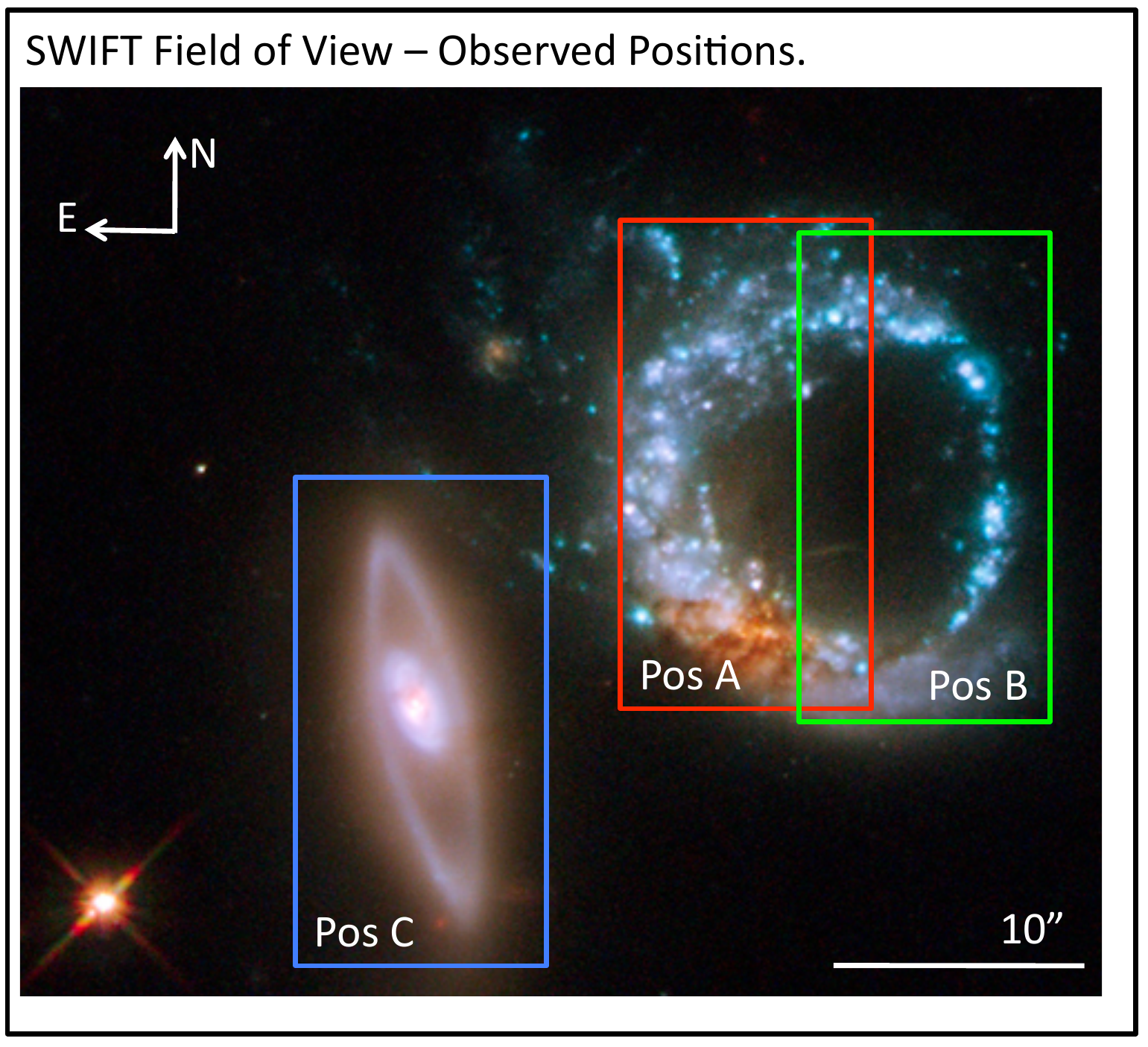}
\caption{A HST/WFPC2 image of the Arp147 system with the observed SWIFT pointings labeled A, B, and C. The image is a false-colour composite of images taken in F450W, F606W, and F814W filters. Credit: NASA, ESA, M.Livio (STScI).}
\label{fig:hst}
\end{figure}

Three pointings, labeled A, B, and C in Figure \ref{fig:hst}, were used to form a mosaic covering the entire system. The central coordinates of the exposures are given in Table \ref{tab:exp} and the field of view is aligned with a PA of 90$^{\circ}$ (i.e. with the long axis of the field of view aligned N-S). To point to the object we first centered the SWIFT field of view on a bright star and offset to the desired position. The coordinates of the offset star are also given in Table \ref{tab:exp}. Two 300s-long exposures were taken at each pointing with an offset sky exposure taken between each object frame, giving an {\tt on-off-on} trio of frames for each pointing. A photometric standard star ($\chi$2 Cet) was observed after the science data was taken. Since this target was part of a bad weather back-up program these data were taken when conditions were less than optimal. During the observations the average seeing was measured from SWIFT exposures of the bright offset star to be $\sim$\,2\farcs{5}. There was also some cirrus cloud interfering with the observations, leading to variations in the sky background between observations. This meant that extra care had to be taken when reducing the data. The reduction procedure is described in Section \ref{sec:red}.

\begin{table}
\begin{tabular}{c|c|c|c}
\hline
Position & RA  & Dec  & Tot. Exp. Time (s) \\
& (h:m:s J2000) & ($^\circ$:':" J2000) & \\
\hline
A & 03:11:18.62 & 01:18:56.8 & 600 \\
B & 03:11:18.08 & 01:18:57.9 & 600 \\
C & 03:11:19.54 & 01:18:48.1 & 600 \\
Star & 03:11:24.81 & 01:16:13.7 & N/A \\
\hline
\end{tabular}
\caption{Coordinates and Exposure Times}
\label{tab:exp}
\end{table}

\subsection{HST/WFPC2 Observations}
Hubble Space Telescope (HST) observations of Arp 147 are available in the Hubble Archives. The system was observed using the Wide Field Planetary Camera 2 (WFPC2) in October 2008, programme number 11902 (P.I. Mario Livio, STScI). Images were taken with the F450W, F606W and F814W filters, which are similar to the B, V and I bands respectively. The data were retrieved from the archive having already been reduced using the STScI pipeline. We then flux-calibrate the images using the PHOTFLAM keyword determined by the pipeline and stitch the exposures together using the IRAF task {\tt wmosaic} from the {\tt stsdas} package.

\section{Data Reduction}
\label{sec:red}

\subsection{SWIFT integral field data}
\label{sec:swift}
We reduce the SWIFT data in several steps using different packages, including the dedicated SWIFT data reduction pipeline (R. Houghton, private communication). Since the instrument has two twin spectrographs (as mentioned in Section \ref{sec:SW}), each with a separate detector, there are two frames to be processed for each SWIFT exposure taken. These correspond to one half of the field of view, split along the short axis, and are labeled {\it master} and {\it slave}. As far as possible we process the master and slave frames separately, though we stitch them together to recover the entire SWIFT field of view at the end of the reduction procedure.

The basic reduction procedure is the same for all science frames (consisting of the object, sky and standard star frames). We prepare the files using a pipeline routine which measures and removes the bias level, trims the overscan from the frame, and flat-fields the data. At this stage we apply the {\tt L.A.COSMIC} routine \citep{vanDokkum01} to identify and flag cosmic rays in the data. The identified cosmic rays are replaced with NaN (i.e. not a number) values.

We find two wavelength solutions (one each for master and slave frames) using Ar and Ne arc lamp exposures. We apply the resulting wavemaps to the prepared two-dimensional science frames, yielding a set of master and slave data cubes. 

We then re-normalise the data cubes along the spectral axis by dividing through with a low-order polynomial representing the flat-field lamp spectrum. We flux calibrate the cubes using the standard star ($\chi$2 Cet) observed on the same night. At this stage we also apply corrections for the differing throughputs of the two spectrograph channels.

\begin{figure}
\includegraphics[width=84mm]{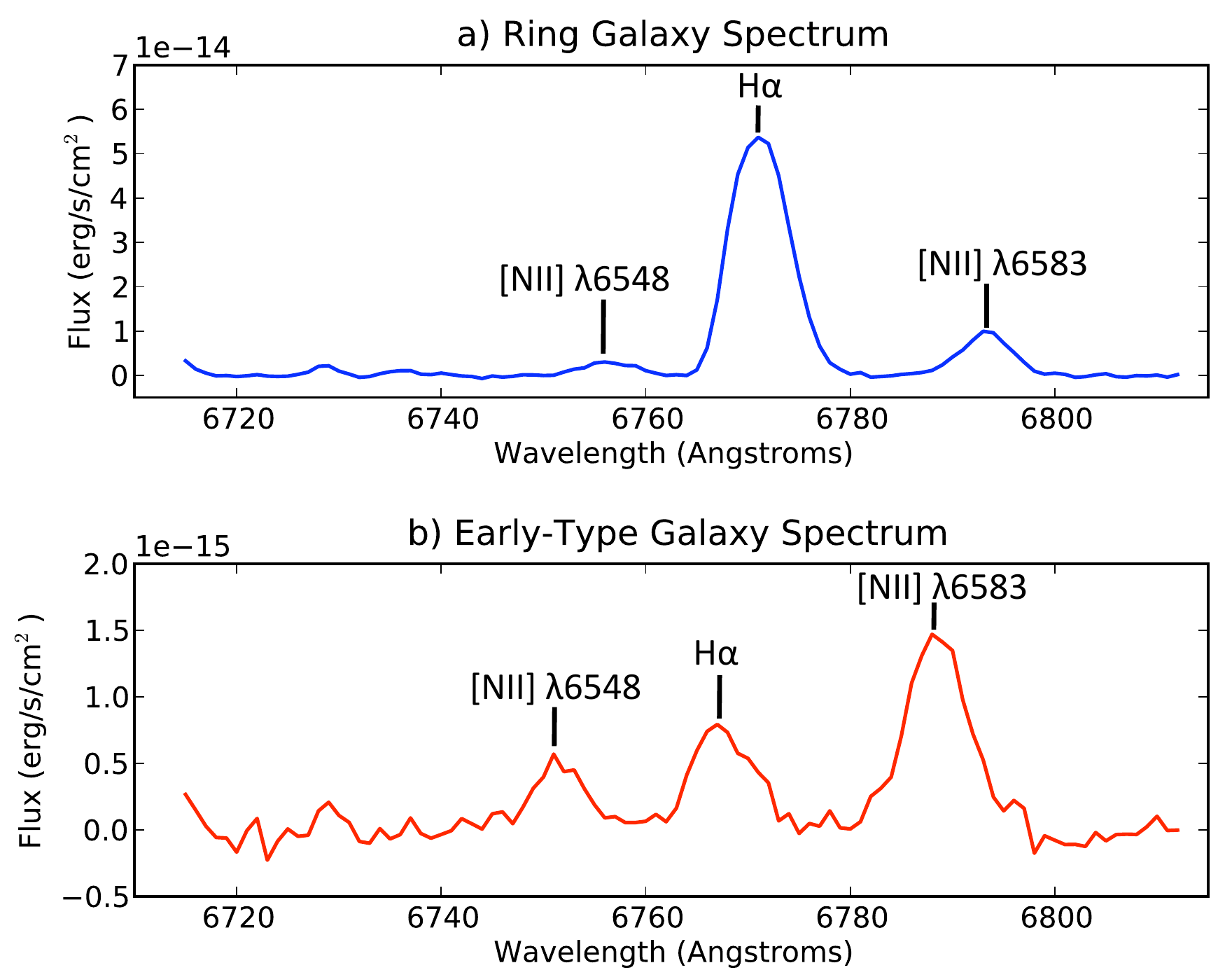}
\caption{Spatially integrated spectra of the ring galaxy (a) and the early-type companion (b) showing the H$\alpha$ and [NII] lines.}
\label{fig:Ha_spec}
\end{figure}

Sky subtraction is complicated by the presence of variable cirrus clouds during the observations, causing large variations in sky brightness on short time scales. We can analyse emission lines that do not require accurate continuum strengths by fitting a low-order polynomial to the galaxy continuum (including any residual sky background). This works well for lines uncontaminated by night sky emission, such as the H$\alpha$ and [NII] lines. We extract spatially integrated spectra in this way and two are shown in Figure \ref{fig:Ha_spec}.

\begin{figure}
\includegraphics[width=84mm]{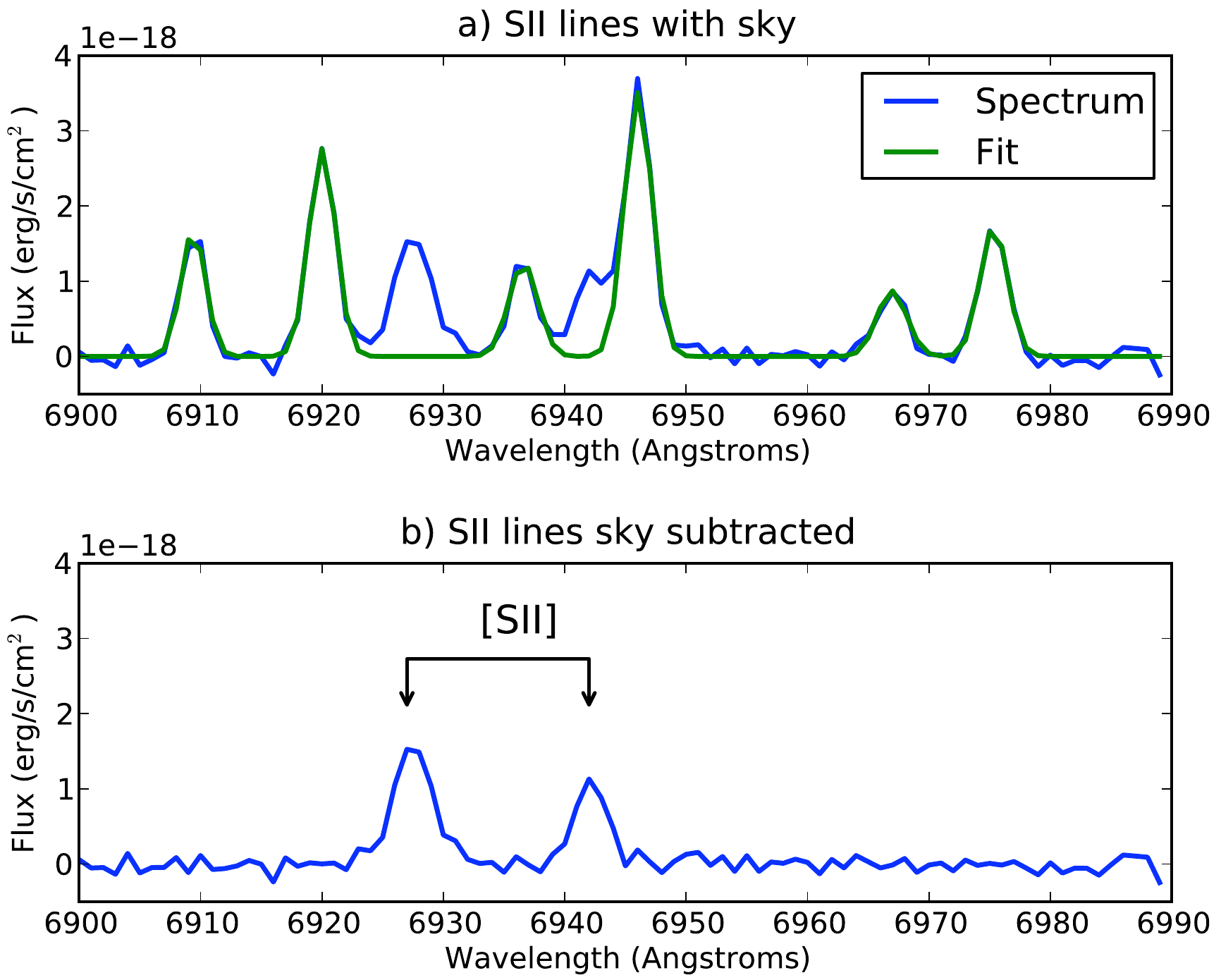}
\caption{The [SII] sky subtraction procedure. Panel a shows the raw spectrum covering the region containing the [SII] emission lines and sky lines, with the Gaussian fits to the sky lines super-imposed. Panel b shows the result of subtracting this fit from the raw spectrum.}
\label{fig:SII}
\end{figure}

The [SII] emission line at 6731~\AA~is contaminated by a sky emission line over a substantial fraction of the ring galaxy. To correctly subtract the sky emission, we fit six sky lines (including the contaminating one) in the near vicinity of the [SII] doublet with Gaussian line profiles. This is done for the {\em sky} spectrum, and the relative strengths, wavelength ratios and line widths derived from the fit are determined. We then hold these relative parameters fixed and fit to the same lines (allowing the overall strength to vary), {\em excluding} the line which contaminates the [SII] emission, in each object spectrum. We use the five other sky lines and the relative parameters already determined from the fit to the sky spectrum to determine the fit parameters for the contaminating sky line. We then subtract this final fit to the sky lines, including the reconstructed contaminating line, from the object spectrum. The results are shown in Figure \ref{fig:SII}.

\subsection{Colours and equivalent widths}
\label{sec:HST}

\begin{figure}
\includegraphics[width=84mm]{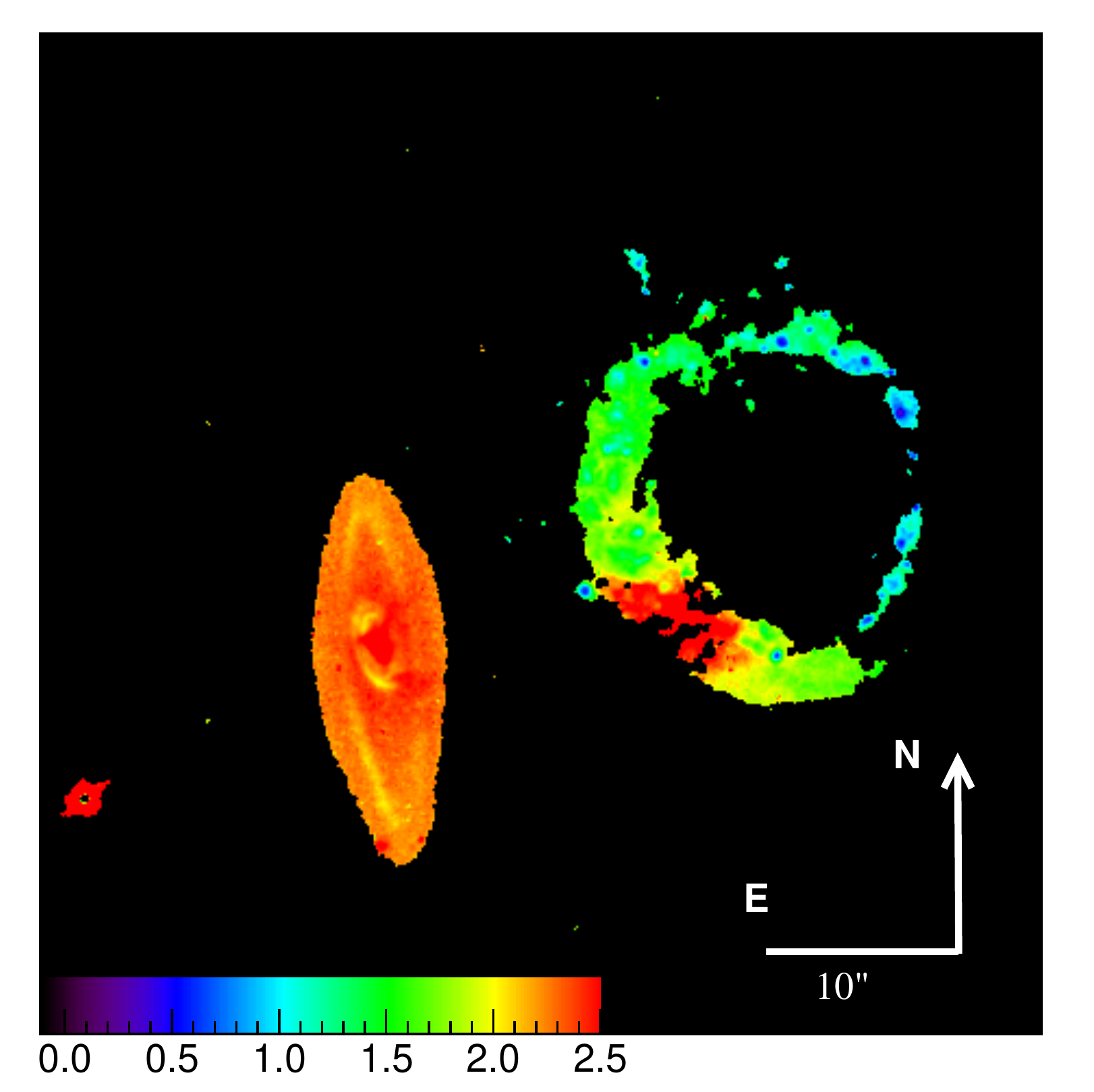}
\caption{F450W$-$F814W colour for the Arp 147 system. The colour values have been corrected to correspond directly with B$-$I colours in the Vega system. The western part of the ring galaxy has significantly bluer colours, while the reddest colours are in the SE region, possibly corresponding to the nuclear bulge of the original disk galaxy from which the ring was formed.}
\label{fig:colour}
\end{figure}

We use the HST data to measure spatially resolved colours for the ring galaxy and the early type galaxy. Multiple exposures are available in each of F450W, F606W, and F814W filters. We combine all exposures in a single filter after re-alignment and rejection of cosmic rays. The resulting data sets are flux calibrated using the latest {\tt PHOTFLAM} values, and registered using world co-ordinate (WCS) information. We then made a mask frame based on a minimum surface brightness cut-off in the F606W image. An F450W$-$F814W (approximately B$-$I) colour map created using the masked, re-aligned, flux-calibrated data sets is shown in Figure \ref{fig:colour} (the error between B$-$I colour and observed F450W$-$F814W colour depends on spectral type, but is typically about 0.02 mag). We also determined the mean colour in each of four quadrants of the ring galaxy (defined in Section \ref{sec:sf}), the measured values are tabulated in Table \ref{tab:EW}. Typical errors on the colours are less than 0.05 mag.

To measure the H$\alpha$ equivalent width it is necessary to have some estimate of the galaxy continuum at wavelengths close to the emission lines. We measure the continuum level from integrated HST photometry in the F606W filter, which has a pivot wavelength of 5919~\AA\, and a bandwidth of 1579~\AA. Given the worse-than-average seeing conditions during our observations, we restricted the spatial resolution of our equivalent width measurements to a single value for each quadrant of the ring galaxy (see Section \ref{sec:sf} and Figure \ref{fig:quad}). We mask the HST image to exclude any sky, and then sum the flux in each quadrant. We then subtract from each quadrant the measured line fluxes in the H$\alpha$ and both [NII] lines in that quadrant, to correctly account for the line contamination. This allows us to compute an estimated continuum level in each quadrant, in units of erg\,s$^{-1}$\,\AA$^{-1}$\,cm$^{-2}$, and the equivalent width for each quadrant.  The same procedure is followed to derive a single integrated value for the early type galaxy.

\section{Results}

\subsection{Collision Geometry}
\label{sec:hakin}

We fit an ellipse to the image of the ring galaxy in the F814W band and use the fitted major and minor axes of the ellipse to find the inclination of the ring using Equation \ref{eqn:i} \citep{Fosbury77}, which assumes an intrinsic circular geometry.

\begin{equation}
\label{eqn:i}
\rm{cos(i)=\frac{minor\ axis}{major\ axis}}
\end{equation}

We find that the inclination of the ring is $25.3\pm1.5^{\circ}$ and the radius of the ring is $5.8\pm0.1\rm{\,kpc}$. The photometric position angle (PA) of the ring is $33\pm4^{\circ}$ where the errors on these values are derived from the fitted ellipse.

\begin{figure}
\includegraphics[width=84mm]{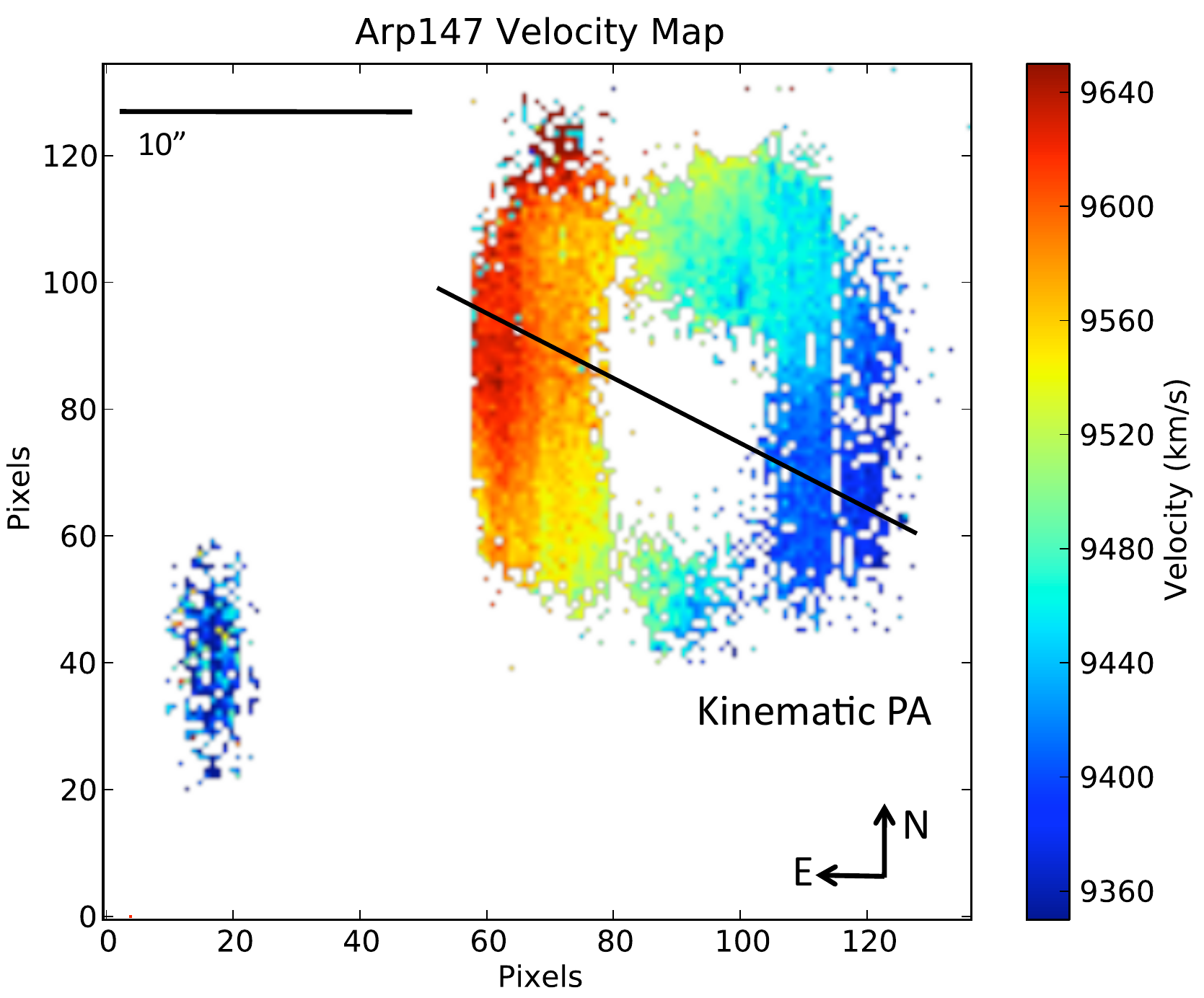}
\caption{The velocity map of the Arp147 system. This is derived by fitting the H$\alpha$ emission line in the ring galaxy with a Gaussian and thus calculating the corresponding line-of-sight velocity shift. The same procedure is used for the early-type galaxy but instead fitting the comparatively stronger [NII]$\lambda6583$ line. The kinematic PA of the ring, measured as described in Section \ref{sec:hakin}, is superimposed.}
\label{fig:kin}
\end{figure}

We obtain a velocity map of the Arp 147 system using the observed emission lines. For the ring galaxy we isolate and fit the H$\alpha$ line with a Gaussian profile in each spaxel where the line has a S/N$\geq$3.5. We then use the centroid of the fit to find the line-of-sight velocity. The procedure is the same for the early-type galaxy, but we use the stronger [NII] emission line. Figure \ref{fig:kin} shows the measured line-of-sight velocity map for the entire system.  

For the early-type galaxy we do not expect a significant line-of-sight velocity gradient, since the emission used to calculate the velocity is not stellar in origin (see Section \ref{sec:sf}). Since we lacked stellar absorption lines to determine the velocity this emission line velocity is included in the map merely for completeness.

\begin{figure}
\includegraphics[width=84mm]{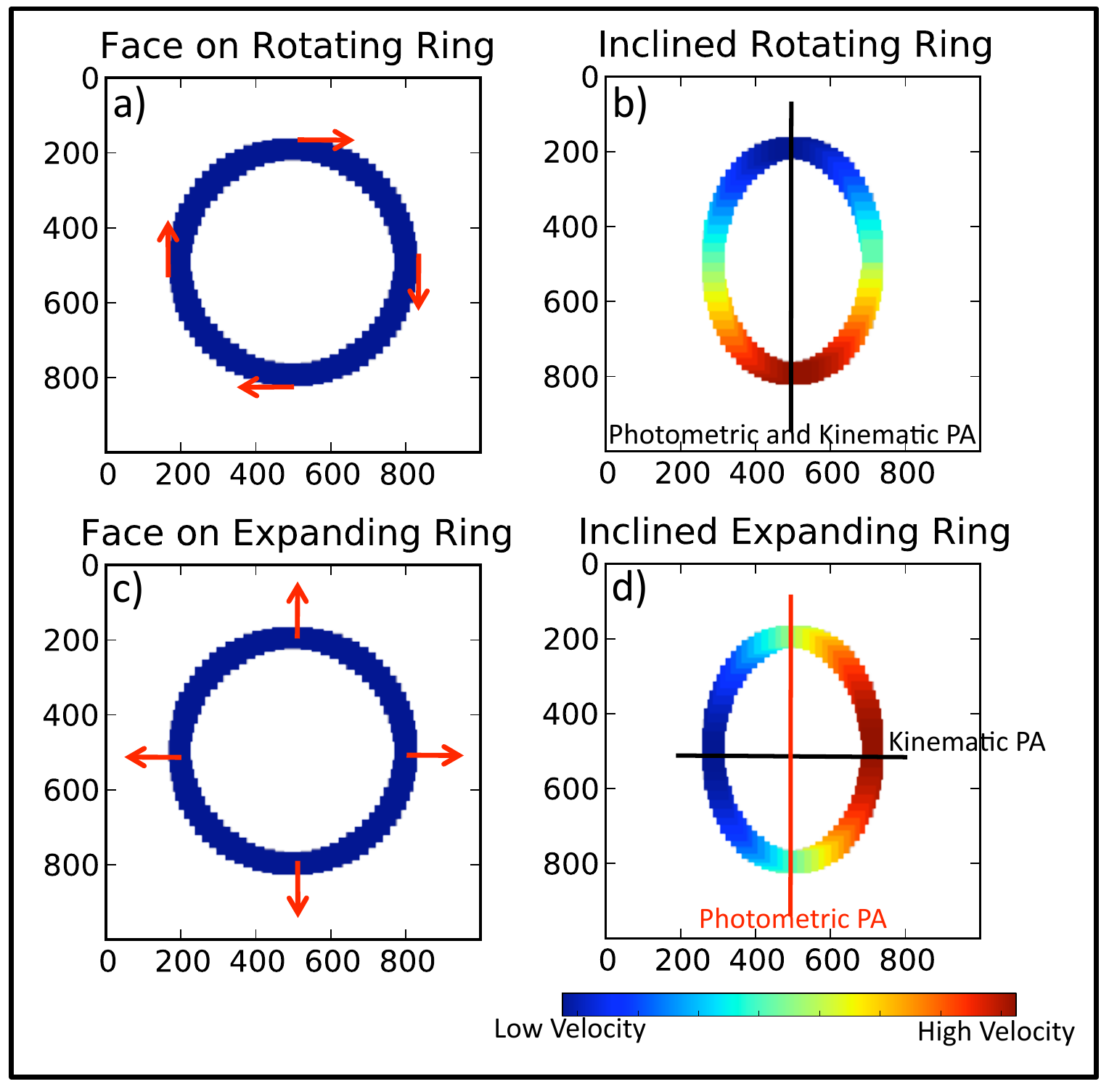}
\caption{The ring galaxy geometry. Panels a and b show that a purely rotating ring will exhibit a kinematic major axis coincident with its photometric major axis. Panels c and d show that a purely expanding ring will exhibit a kinematic major axis perpendicular to its photometric major axis.}
\label{fig:geometry}
\end{figure}

The observed line-of-sight velocity gradient across the ring galaxy is due to rotation and expansion of the ring. These two components must be disentangled from each other. As shown in Figure \ref{fig:geometry} we do this by measuring the photometric and kinematic position angles (PAs) of the ring. Assuming an intrinsic circular geometry, if the ring is purely rotating we expect to measure the kinematic and photometric PAs to be the same. This is illustrated in Figures \ref{fig:geometry}a and \ref{fig:geometry}b. If the ring is purely expanding we expect to measure the kinematic and photometric PAs to be perpendicular to each other. This is illustrated in Figures \ref{fig:geometry}c and \ref{fig:geometry}d. Therefore if we calculate the angle between the measured kinematic and photometric PAs of the ring galaxy it is possible to calculate how much of the measured line-of-sight velocity is in rotation, and how much is in expansion. We measure the kinematic PA of the ring galaxy as $70.5\pm0.5^{\circ}$ using an IDL routine (by M. Cappellari, private communication) which uses the method described in Appendix C of \citet{Krajnovic06}. The kinematic PA is shown superimposed on the velocity map in Figure \ref{fig:kin}.

To measure the component of the observed velocity due to expansion we plot each measured velocity point as a function of the azimuthal angle of the ring, $\theta$. We then fit the resulting graph with Equation \ref{eqn:exp_vel} \citep{Fosbury77}, where $\rm{V_{syst}}$ is the systemic velocity of the ring, $\rm{V_{rot}}$ the component of the observed velocity due to rotation and $\rm{V_{exp}}$ the component due to expansion of the ring.

\begin{equation}
\label{eqn:exp_vel}
\rm{V=V_{syst}+sin(i)(V_{exp}sin(\theta)+V_{rot}cos(\theta))}
\end{equation}

The deprojected edge-to-edge expansion of the ring, $\rm{V_{exp}}$, is $225\pm8\rm{\,km\,s^{-1}}$, giving a radial expansion velocity of $113\pm4\rm{\,km\,s^{-1}}$. Given the radius of the ring ($5.8\pm0.1\rm{\,kpc}$) and assuming a uniform expansion velocity from a point at the centre of the ring, this implies that the collision occurred a maximum of $50\rm{\,Myrs}$ ago. Comparatively, there is little rotational velocity, with a value for $\rm{V_{rot}}$ of $47\pm8\rm{\,km\,s^{-1}}$.

We also measure the systemic velocities of the ring and the early-type companion by fitting emission lines in integrated spectra of each object. The difference between the line-of-sight velocity of the two objects is $151\pm7\rm{\,km\,s^{-1}}$. Given the upper limit on the collision timescale found above we use the age of the minimum age of the interaction-triggered young population, taken to be 1.1\,Myrs (see Section \ref{sec:history} and Table \ref{tab:EW}) to calculate a lower limit for the interaction timescale. 

Since the star formation was more than likely delayed from the time of impact \citep{Mihos94} we arbitrarily assume it took a minimum of $10\rm{\,Myrs}$ for star formation to be triggered \citep{Rappaport10} and use $11.1\rm{\,Myrs}$ as a lower limit on the interaction timescale. This implies that the early-type companion has travelled between $1.7\pm0.1\rm{\,kpc}$ and $7.8\pm0.5\rm{\,kpc}$ along the line of sight since the collision occurred. The distance between the centres of the galaxies in the plane of the sky is $12.8\pm0.1\rm{\,kpc}$ which implies that the collision occured at an impact angle of between $33^{\circ}$ and $54^{\circ}$. A $0^{\circ}$ impact angle implies a head-on collision. 

\begin{table}
\begin{tabular}{cc}
\hline
Property & Value \\
\hline
Ring Inclination & $25.3\pm1.5^{\circ}$ \\
Ring Photometric PA & $33\pm4^{\circ}$ \\
Ring Kinematic PA & $70.5\pm0.5^{\circ}$ \\
Ring Radius & 5.8$\pm0.1\rm{\,kpc}$ \\
Ring Expansion Velocity & $225\pm8\rm{\,km\,s^{-1}}$ \\ 
On Sky Separation$^{a}$ & $12.8\rm{\,kpc}$ \\
Maximum Interaction Timescale$^{b}$ & $50\rm{\,Myrs}$ \\
Impact Angle Range & $33^{\circ}-\,54^{\circ}$ \\
\hline
\end{tabular}
\caption{Kinematic Properties of the Arp 147 System. $^{a}$On sky separation between the ring and companion. $^{b}$ Upper limit on interaction timescale as described in the text.}
\label{tab:kin_props}
\end{table}

\subsection{Star Formation}
\label{sec:sf}

\begin{figure}
\includegraphics[width=84mm]{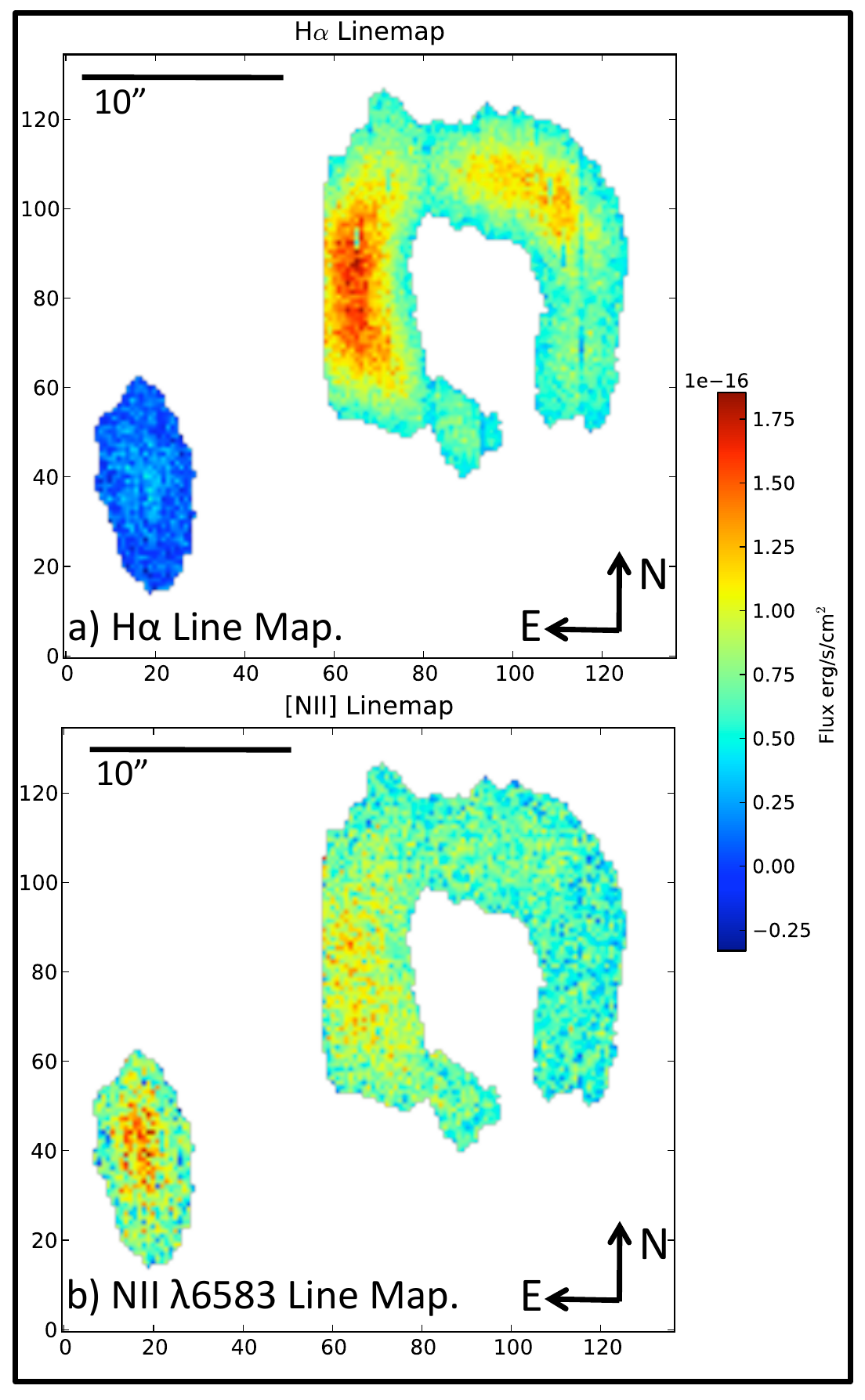}
\caption{a) The H$\alpha$ linemap of the Arp147 system. b) The [NII] $\lambda6583$ linemap of the Arp147 system. Both maps have been continuum subtracted.}
\label{fig:linemaps}
\end{figure}

\begin{figure}
\includegraphics[width=84mm]{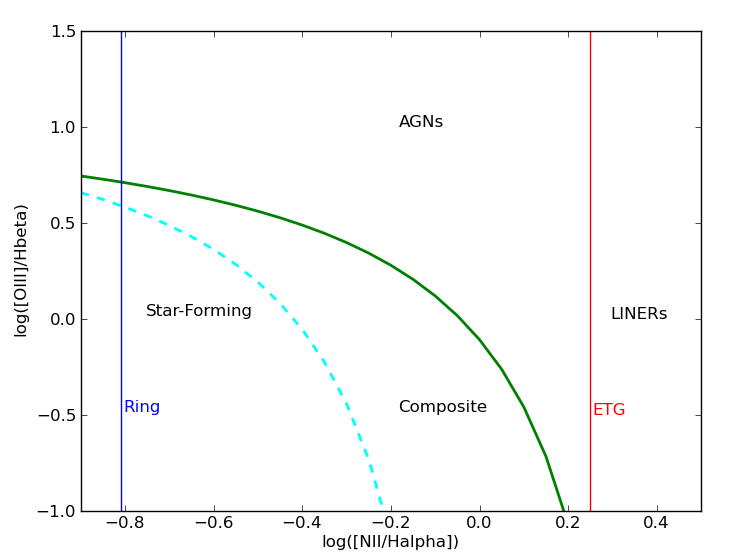}
\caption{A BPT diagnostic diagram showing the 'extreme starburst line' (green ref) and the empirical line below which all galaxies are starbusts (dashed cyan ref). The red line vertical line represents the early-type galaxy, consistent with a non-star-forming ionisation mechanism (either a LINER or a Seyfert galaxy) and the blue line represents the star-forming ring galaxy}
\label{fig:bpt}
\end{figure}

The SWIFT H$\alpha$ and [NII] $\lambda6583$ linemaps for the Arp147 system are shown in Figures \ref{fig:linemaps}a and \ref{fig:linemaps}b respectively. The maps are continuum subtracted and the wavelength ranges used are centred on the lines in question (with different ranges for each object accounting for differences in line-of-sight velocity). The ring galaxy is obviously strongly star-forming. The source of ionisation in the early-type companion is less obvious, however. Typically emission line objects are classified by comparison of their line ratios through the use of so-called BPT diagrams \cite{BPT81}. Figure \ref{fig:bpt} shows the parameter space of a typical BPT diagram which compares the log([NII]/H$\alpha$) and log([OIII]/H$\beta$) ratios. Our spectra do not enough lines to place the objects in the Arp147 system onto a BPT diagram, but it is possible to distinguish the mechanisms behind the line emission in these objects none the less.

\cite{Kewley01} used stellar evolution models to derive a theoretical limit for starburst galaxies on the BPT diagram. This line is shown as the green curve in Figure \ref{fig:bpt} and represents an extreme upper limit for starburst galaxies. \cite{Kauffmann03} revised this downwards to separate AGN and star-forming galaxies in an empirical way. Their line is shown in Figure \ref{fig:bpt} as a dashed cyan line. Objects between the two lines are considered to be ``composite'' objects whose ionising spectrum arises due to a combination of star formation and AGN. For the Arp 147 galaxies we can measure only one parameter of the BPT diagram - log([NII]/H$\alpha$). The ring galaxy is shown as a blue line along its possible log([OIII]/H$\beta$) values. It is clear that this object is strongly star-forming (not that there was any doubt). The early-type companion is shown as a red line covering its possible log([OIII]/H$\beta$) values. This galaxy clearly falls in the LINER/Seyfert part of the BPT diagram. 

\begin{figure}
\includegraphics[width=84mm]{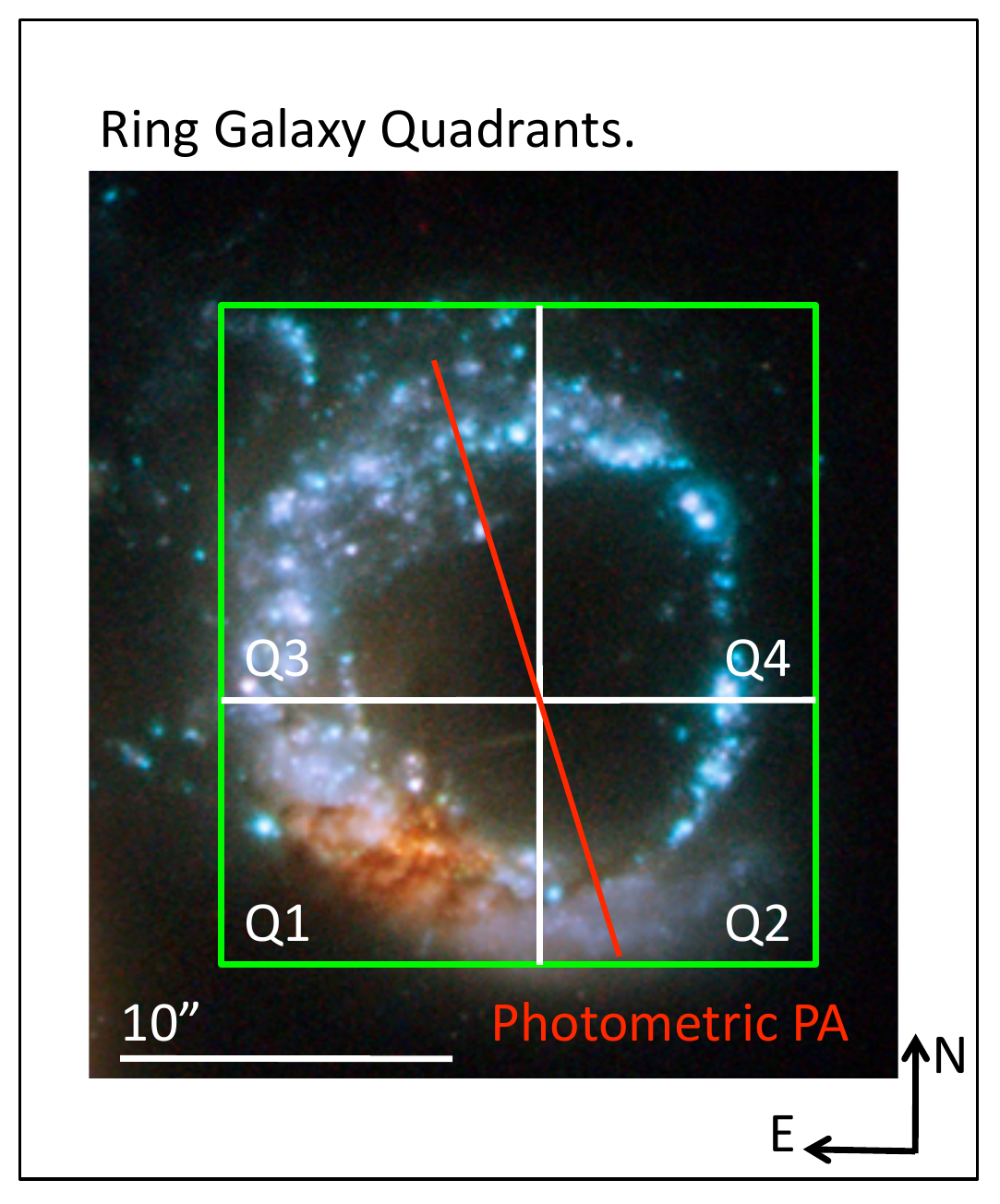}
\caption{The ring galaxy was split into the four quadrants shown, labelled 1, 2, 3 and 4. Quadrant 1 was chosen to include the redder area in the south-east corner, thought to correspond to the nuclear bulge of the original disk galaxy. A spatially integrated spectrum was extracted for each quadrant.}
\label{fig:quad}
\end{figure}

We extract spatially integrated spectra for each of the two galaxies in the system. We also study the ring in more detail to see if any of the physical properties vary over its extent - we extract four spectra from quadrants across the ring (see Figure \ref{fig:quad}). We chose this approach to ensure the spatially integrated spectra had enough S/N to measure the weaker emission lines. The quadrants are not equally sized and were chosen such that quadrant 1 covered the redder region to the south-east of the ring. It is possible that this region corresponds to the nuclear bulge of the original disk galaxy from which the ring was formed.

We measure the integrated H$\alpha$ flux for the ring galaxy and use this to derive the star formation rate according to Equation \ref{eq:SFR} \citep{Kennicutt98}. We measure a star formation rate of 4.68$\pm$0.67\,M$_{\odot}$/yr, as shown in Table \ref{tab:Ha}.

\begin{equation}
\label{eq:SFR}
\rm{SFR(M_{\odot}yr^{-1})=7.9\times10^{-42}L(H\alpha)(erg\,s^{-1})}
\end{equation}

\subsubsection{Star Formaton history}
\label{sec:history}
We measure the H$\alpha$ equivalent width for each quadrant of the ring galaxy (see Section \ref{sec:HST}), using spatially integrated spectra from each region. The results are presented in Table \ref{tab:EW}. Using the measured equivalent widths and colours, we can constrain the age range of the recent star formation activity occurring in the ring galaxy, by comparing with stellar population synthesis models. In particular, we use the Starburst99 \citep{Starburst99} code, with a standard \citet{Salpeter55} initial mass function (IMF) with upper and lower mass cutoffs of $0.1\rm{\,M_{\odot}}$ and $100\rm{\,M_{\odot}}$ respectively, and solar metallicity.  We consider both instantaneous starbursts and continuous star formation as the two limiting cases for the young (encounter triggered) stellar population. This is consistent with \citet{Mihos94}, who predict an enhancement in star formation in the ring of a factor of $2-3$. 

Our analysis to derive the age range of the recent star formation is best understood as a step-by-step process.

\begin{figure}
\includegraphics[width=84mm]{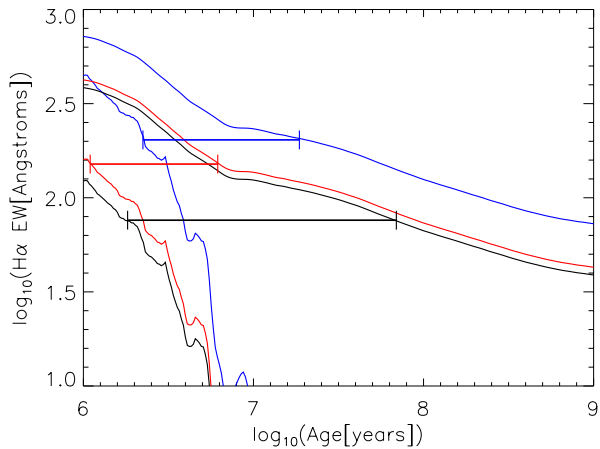}
\caption{Plot showing the range of allowable ages for the recent star formation in three of the four quadrants of the ring galaxy. There are two curves for each quadrant (black for Q2, red for Q3 and blue for Q4), representing the extremes of the allowed values for the H$\alpha$ EW as a function of age of the young (encounter triggered) stellar population.  These extremes take into account instantaneous and continuous star formation histories, and all allowable ages for the older (pre-encounter) stellar population in the disk galaxy. The observed value of the H$\alpha$ EW in each quadrant defines an age range for the young stars, these are shown by horizontal bars for each quadrant.}
\label{fig:age}
\end{figure}

{\em Step 0: Assumptions} It is reasonable to expect that the observed light from the ring galaxy consists of two components, one stemming from an old population (native stellar population of the disk galaxy prior to the encounter), and a second younger population, triggered by the encounter.  We {\em assume} that the old population native to the disk galaxy has the same age in all quadrants.  Also, we {\em assume} negligible extinction in all quadrants of the ring. The latter assumption is based on the visual appearance of the ring (see Figure \ref{fig:hst}) and the colour map (Figure \ref{fig:colour}), the low value of the inclination, and is partly motivated by the lack of sufficient data to derive a reliable extinction estimate. The reddest quadrant (Q1 of Figure \ref{fig:quad}) which appears to contain the nuclear bulge of the original disk galaxy may contain a stellar population of a different age (bulge component), and is treated separately. 

{\em Step 1: Constraining the age range of the old stellar population} The reddest B$-$I colour we observe in a non-nuclear quadrant (Q2) is 1.6, so the age of the old component of the disk galaxy is restricted to correspond to B$-$I colour redder than 1.6 (the colour gets bluer when we add a young component). That gives a lower bound on the age of the old component of 1.3 Gyr, the upper bound is the age of the Universe, or 13 Gyr. (We note that the B$-$I colour of the neighbouring early type galaxy is consistent with a stellar population of 13 Gyr age).  Knowing the age (and hence colour) of the old component, we can determine the fraction of light from young stars. The observed H$\alpha$ equivalent width (EW) in each quadrant places a further constraint on the old component colour, as too small a fraction of young stars cannot simultaneously account for the observed H$\alpha$ EW and B$-$I colour. This increases the lower limit on the old component to 1.8 Gyr, if the young stars form in an instantaneous starburst, or 1.5 Gyr if they form continuously.

{\em Step 2: Fraction of light from young stars} For a fixed old component age, we can compute the young population fraction required to produce the observed colour in each quadrant, as a function of age of the young component. Two curves are computed, one for the young stars forming in a single burst (instantaneous) or forming continuously.  Varying the old component age over the allowed range gives us two families of curves showing the fraction of light from young stars, one for instantaneous and the other for continuous star formation.

{\em Step 3: Computing the H$\alpha$ EW v/s young component age} Each curve of light fraction v/s age of the young component (for a fixed old population age) corresponds to a curve of observed H$\alpha$ EW v/s age of young component.  Figure \ref{fig:age} show the most extreme curves for three quadrants (excluding the nuclear quadrant) -- black for Q2, red for Q3, and blue for Q4. 

{\em Step 4: Constrain age and light fraction using H$\alpha$ EW} For an observed H$\alpha$ EW, the region between the most extreme curves corresponds to the allowed age range for the young component in each quadrant, this is shown by horizontal bars in Figure \ref{fig:age}.  We get an age range, rather than a fixed value, due to the uncertainty in knowing the age of the old component, and the recent star formation history (i.e. instantaneous or continuous). For each quadrant, we can derive an allowed age range and the corresponding fraction of light from the young star formation triggered by the encounter. These are presented in Table \ref{tab:EW}. The age range for the young stars is modest, ranging between 1.1 and 70 Myr, independent of star formation history. We use the mass-to-light ratio for each component provided by the Starburst99 models to also derive a mass fraction in young stars.

It is interesting to note that although the young stars produce as much as a third of the V band light in the ring galaxy Q4, the mass fraction is negligible at $\sim$1\% in all cases, irrespective of which recent star formation mode we adopt -- a robust result. We also note that all quadrants, including those dominated by blue light from young stars {\em require} a two component stellar population to fit the measured values for the colour and equivalent width simultaneously. 

In quadrant 1 of the ring galaxy, the observed colour is too red (B$-$I = 2.09) to allow a significant age range for the old population. Indeed, the visual appearance (see Figure \ref{fig:quad}) suggests that some reddening may be due to dust.  For consistency with the other measurements, we compute an age range for Q1, assuming that the old population is 13 Gyr old. The results are shown in Table \ref{tab:EW}. The corresponding light fraction for the young component is very small, and the mass fraction is entirely negligible, as expected.

We do not do this computation for the early type galaxy, as the observed H$\alpha$ emission has a different origin, as discussed in Section \ref{sec:sf} above.

\subsubsection{Total stellar mass}
\label{sec:mass}
Using the results of the previous section for the light fraction from the old component in each quadrant of the ring galaxy, and using the total observed F606W continuum flux from the ring galaxy, we compute the total stellar mass of the ring galaxy to be in the range $6.1-10.1 \times 10^{10}\,M_{\odot}$, the uncertainty stemming from the age of the old stellar component that dominates the mass. Thus, it appears as though the ring galaxy is a typical disk galaxy, similar to the Milky Way. The ``nuclear bulge'' (Q1) comprises 30-50\% of the total mass.

\begin{table}
\begin{tabular}{ccc}
\hline
Object & L(H$\alpha$) (erg/s)& SFR (M$_{\odot}$/yr) \\
\hline
Ring & $5.92(\pm0.85)\times10^{41}$ & $4.68\pm0.67$ \\
\hline
\end{tabular}
\caption{The H$\alpha$ flux and star formation rate for the ring galaxy.}
\label{tab:Ha}
\end{table}

\begin{table*}
\begin{tabular}{cccccc}
\hline
Object & EW(H$\alpha$)& B$-$I & $\log_{10}{\rm Age}$ & Light & Mass\\
 & (\AA) & colour & (years) & fraction & fraction\\
\hline
Ring Q1& $-83\pm2.5$ & 2.09 & 6.06 -- 6.27  & $\sim$0.04 & -- \\
Ring Q2 & $-76\pm1.6$ & 1.60 & 6.26 -- 7.84 & 0.05 -- 0.2 & $<$1\%\\
Ring Q3 & $-151\pm3.0$ & 1.55 & 6.04 -- 6.79 &  0.06 -- 0.13 & $<$0.1\%\\
Ring Q4 & $-203\pm5.8$ & 1.21 & 6.35 -- 7.27 & 0.14 -- 0.34 & $<$0.15\%\\
ETG & $-4.1\pm0.3$ & 2.27 & -- & -- & -- \\
\hline
\end{tabular}
\caption{The measured H$\alpha$ equivalent widths, young starburst age ranges, light and mass fractions for the quadrants of the ring galaxy.}
\label{tab:EW}
\end{table*}

\begin{table*}
\begin{tabular}{c|c|c|c|c}
\hline
Object & 12+\rm{log(O/H)} & S2 & n$_{e}$ & log(u)\\
\hline

Ring & $8.44\pm0.04$ & $1.16\pm0.32$ & 201cm$^{-3}$ & $-3.1\pm0.1$ \\
Quad1 & $8.50\pm0.07$ & $1.16\pm0.15$ & 297cm$^{-3}$ & $-2.9\pm0.1$ \\
Quad2 & $8.28\pm0.24$ & $1.25\pm0.21$ & 201cm$^{-3}$ & $-3.0\pm0.1$ \\ 
Quad3 & $8.52\pm0.07$ & $1.20\pm0.58^a$ & 267cm$^{-3}$ & $-3.1\pm0.1$ \\
Quad4 & $8.19\pm0.13$ & $1.25\pm0.20$ & 153cm$^{-3}$ & $-2.8\pm0.1$ \\

\hline
\end{tabular}
\caption{The $S2$ parameter, electron density and ionisation parameter. $^a$ The [SII] $\lambda6731$ line in this quadrant was particularly badly contaminated by sky so it was necessary to fix the width of the line to match that of the [SII] $\lambda6716$, resulting in a large error on the fit to the line.}
\label{tab:S2lines}
\end{table*}

\subsection{Line Ratios and Derived Properties.}
The N2 ratio ($\rm{N2}=log(\rm{[NII]/H\alpha})$) is a relatively reliable empirical indicator of oxygen over-abundance \citep{PettiniPagel04}. This relation is justifiable because at the metallicity range studied here oxygen can be assumed to be a product of primary nucleosynthesis and nitrogen a product of secondary nucleosynthesis. Therefore the strength of nitrogen emission gives an indication of the over-abundance of oxygen in the star-forming region \citep{Storchi94}. We calculate oxygen overabundance for the ring galaxy and for each quadrant of the ring using Equation \ref{eq:OH}, where the solar value is 12+log(O/H)=8.76$\pm$0.07 \citep{Caffau08}. The results are presented in Table \ref{tab:S2lines}. Of note is the fact that the eastern half of the ring shows a roughly solar oxygen abundance, whereas the western half of the ring shows an oxygen abundance a factor of two lower.

\begin{equation}
\label{eq:OH}
\rm{12+log(O/H)=8.9+0.57\cdot log([NII]/H\alpha)}
\end{equation}

Since the [SII] $\lambda\lambda$ 6716, 6731 emission lines have similar excitation energy but differing collisional de-excitation rates, the observed emission strengths of these two lines are dependent on the electron density in the emitting region \citep{Osterbrock89}. We use the ratio $\rm{S2}=[SII](\lambda6716)/[SII](\lambda6731)$ to calculate the electron density, n$_{e}$, in each quadrant of the ring galaxy and in the early-type companion. We use the standard iraf {\tt nebular} package to find the values which are shown in Table \ref{tab:S2lines}. All quadrants of the ring galaxy have typical electron densities expected for star-forming HII regions \citep{KKB89}.

The ionisation parameter, u, is the ratio of the ionising photon density to the particle density in the HII region and is one of the parameters defining the emission line spectra of HII regions. It can be found using the measured ratio $\rm{[SII](6717+6731)/[SIII](9069+9532)}$ and Equation \ref{eq:U} (see \citet{Diaz91}). The value of log(u) was calculated for the ring galaxy and each quadrant of the ring. The calculated values of log(u) are shown in Table \ref{tab:S2lines}. 

\begin{equation}
\label{eq:U}
\rm{log(u)=-1.684\cdot log([SII]/[SIII])-2.986}
\end{equation}

The ionisation parameter is expected to fall with increasing metallicity \citep{EvansDop85}, and our measurements roughly follow this expected trend with oxygen abundance.

\section{Conclusions and Discussions}

The ring galaxy in the Arp 147 system is expanding with a radial velocity of $113\pm4\,\rm{kms^{-1}}$. Assuming the expansion is a result of the impact which originally formed the ring, and assuming the expansion velocity remained constant and all material in the ring originated at a point in the centre of the ring, this implies an upper limit on the timescale for the collision of 50\,Myrs. The relative velocity between the early-type galaxy and the ring is $151\pm7\,\rm{kms^{-1}}$. 

\citet{Mihos94} have created models of collisionally created ring galaxies by studying centred head-on collisions. Using the measured kinematics of Arp 147 it is shown that the impact which created Arp 147 probably occurred at an angle in the range $33^{\circ}-54^{\circ}$. Evidence presented by \citet{Gerber92} implies that an off-centre collision better produces an empty ring with a dislodged nucleus, like that seen in Arp 147. Thus we speculate that an off-centre, angled collision created the Arp 147 system. The parameters we derive for the impact are shown in Table \ref{tab:kin_props} and could be used to constrain future models. In their simulation \citet{Gerber92} found that the nuclear region of the original disk was not only dislodged during the collision but was offset from the plane of the resulting ring. They identified the dislodged nucleus with quadrant 1 of the actual ring as we do, but we see no evidence in our kinematics to suggest that the nucleus is indeed offset from the plane of the ring.

We measure the equivalent width of the H$\alpha$ lines in the ring galaxy, and the B$-$I colours from multi-band HST photometry.  Comparing this with the results from the Starburst99 stellar population synthesis models \citep{Starburst99}, we conclude that we require a two component stellar population throughout the ring galaxy, even in the bluest quadrant whose visual appearance is dominated by light from young stars. We constrain the age range of the recent star formation activity in the ring, finding that the young stellar population of the ring galaxy is between 1.1 -- 70 Myr old, with the most extreme assumptions for star formation history (instantaneous and continuous).  We find the corresponding fraction of light from the young stars to span the range 0.04 to 0.34, while the stellar mass formed during the recent star formation activity is negligibly small, below 1\% and likely $\sim$0.1\%. 

The models of \citet{Mihos94} give a good description of the evolution of star formation in the created ring in the aftermath of the collision. They find that some time after the encounter star formation tends to be dominated by the created ring. Their simulations show that the created ring moves outwards, sweeping up gas and triggering star formation. The inner parts of the ring are left devoid of star formation. The star formation in the ring is therefore not triggered immediately upon collision with the companion, but with a certain delay, which they estimate to be $\sim 10\,\rm{Myrs}$.  In the case of Arp 147, where the encounter was likely off-centre and not head-on, the colour map suggests an age gradient in the recent star formation, with the bluest colours in the west-southwest region, and progressively getting redder counterclockwise. This observation is consistent with the scenario that the ``ring'' galaxy is not a planar configuration, but is actually a ``corkscrew'' shape that appears planar when projected along the line-of-sight (M. Livio, priv. comm). Further detailed modelling will be required to confirm or deny this hypothesis.

The redder colours in quadrant 1 of the ring could be due to dust reddening or they may indicate that this region corresponds to the nuclear bulge of the disk galaxy from which the ring was created. We compute a total stellar mass of $6.1-10.1 \times 10^{10}\,M_{\odot}$ for the ring galaxy, making it a rather typical disk galaxy prior to the encounter. We observe that between 30--50\% of the stellar mass is in Q1, lending weight to the interpretation that this quadrant contains the bulge of the disk galaxy. 

We measure the oxygen over-abundance for the ring galaxy and in the four quadrants of the ring. The eastern half of the ring has an oxygen abundance a factor of $\sim$\,2 higher than that of the western half of the ring. The electron density and ionisation parameter values for the quadrants of the ring are consistent with those expected in star-forming HII regions.

\section*{Acknowledgments}
The Oxford SWIFT integral field spectrograph is directly supported by a Marie Curie Excellence Grant from the European Commission (MEXT-CT-2003-002792, Team Leader: N. Thatte). It is also supported by additional funds from the University of Oxford Physics Department and the John Fell OUP Research Fund.  Additional funds to host and support SWIFT at the 200-inch Hale Telescope on Palomar are provided by Caltech Optical Observatories.

Based on observations obtained at the Hale Telescope, Palomar Observatory, as part of a collaborative agreement between the California Institute of Technology, its divisions Caltech Optical Observatories and the Jet Propulsion Laboratory (operated for NASA), and Cornell University.

Based on observations made with the NASA/ESA Hubble Space Telescope, and obtained from the Hubble Legacy Archive, which is a collaboration between the Space Telescope Science Institute (STScI/NASA), the Space Telescope European Coordinating Facility (ST-ECF/ESA) and the Canadian Astronomy Data Centre (CADC/NRC/CSA)

Some of the data presented in this paper were obtained from the Multimission Archive at the Space Telescope Science Institute (MAST). STScI is operated by the Association of Universities for Research in Astronomy, Inc., under NASA contract NAS5-26555. Support for MAST for non-HST data is provided by the NASA Office of Space Science via grant NNX09AF08G and by other grants and contracts.

L. Fogarty would like to thank Prof. M. Livio (STScI) for interesting discussions regarding the ring geometry, and acknowledge the generous support of the Foley-B\'ejar Scholarship through Balliol College, Oxford and the support of the STFC.

\bibliographystyle{mn2e}
\bibliography{Arp147}
\end{document}